\documentclass[11pt,preprint]{aastex}
\usepackage{graphicx}
\usepackage{graphics}
\usepackage{color}
\usepackage{epsfig}
\usepackage{epsf}

\def\chandra{{\sl Chandra\ }}
\def\xmm{{\sl XMM-Newton\ }}

\shortauthors{}
\shorttitle{}
\begin{document}

\title{{\sl Chandra} Observation of the Edge-on Spiral NGC 5775:
Probing the Hot Galactic Disk/Halo Connection}

\author{Jiang-Tao Li\altaffilmark{1,2},
Zhiyuan Li\altaffilmark{2}, Q. Daniel Wang\altaffilmark{2}, Judith
A. Irwin\altaffilmark{3}, \& Joern Rossa\altaffilmark{4} }
\altaffiltext{1}{Department of Astronomy, Nanjing University, 22
Hankou Road, Nanjing 210093, P. R. China}
\altaffiltext{2}{Department of Astronomy, University of
Massachusetts, 710 North Pleasant Street, Amherst, MA 01003, U.S.A.}
\altaffiltext{3}{Department of Physics, Engineering Physics \&
Astronomy, Queen's University, Kingston, ON K7L 3N6, Canada}
\altaffiltext{4}{Department of Astronomy, University of Florida, 211
Bryant Space Science Center, Gainesville, FL 32611, U.S.A.}

\begin{abstract}
We study the edge-on galaxy NGC~5775, utilizing a 58.2 ks {\sl
Chandra} ACIS-S observation together with complementary {\sl HST}
ACS, {\sl Spitzer} IRAC and other multi-wavelength data sets. This
edge-on galaxy, with its disk-wide active star formation, is
particularly well-suited for studying the disk/halo interaction on
sub-galactic scales. We detect 27 discrete X-ray sources within the
$D_{25}$ region of the galaxy, including an ultra-luminous source
with a 0.3-7 keV luminosity of $\sim7\times10^{40}\rm~ergs~s^{-1}$.
The source-removed diffuse X-ray emission shows several prominent
extraplanar features, including a $\sim10\rm~kpc$ diameter
``shell-like'' feature and a ``blob'' reaching a projected distance
of $\sim25\rm~kpc$ from the galactic disk. The bulk of the X-ray
emission in the halo has a scale height of $\sim$1.5 kpc and can be
characterized by a two-temperature optically thin thermal plasma
with temperatures of $\sim$ 0.2 and 0.6 keV and a total 0.3-2 keV
luminosity of $\sim3.5\times10^{39}\rm~ergs~s^{-1}$. The
high-resolution, multi-wavelength data reveal the presence of
several extraplanar features around the disk, which appear to be
associated with the in-disk star formation. We suggest that hot gas
produced with different levels of mass loading can have different
temperatures, which may explain the characteristic temperatures of
hot gas in the halo. We have obtained a sub-galactic scale
X-ray-intensity-star formation relation, which is consistent with
the integrated version in other star forming galaxies.
\end{abstract}


\section{Introduction}

It is generally believed that extraplanar diffuse X-ray emission
detected around actively star-forming disk galaxies arises primarily
from the so-called disk/halo interaction (Cox 2005 and references
therein; Wang et al. 1995, 2001, 2003; Strickland et al. 2004a,b).
Existing observational work has concentrated on the overall
correlation between the extraplanar X-ray emission and the
integrated properties of various tracers of star formation in such
galaxies, especially those with nuclear starbursts. Systematic
studies with {\sl Chandra} data (Strickland et al.~2004a,b) have
revealed the correlation between the presence of extraplanar hot gas
and warm ionized gas, which is interpreted as a result of the
interaction between star formation-driven outflow of hot gas and the
cooler ambient interstellar medium (ISM). The integrated extraplanar
X-ray luminosity is found to be quasi-linearly correlated with the
star formation rate, implying that the mechanical energy release
from supernova explosions (SNe) and massive stellar winds is
converted into radiative energy with a certain efficiency. Similar
correlations have been demonstrated in an {\sl XMM-Newton} survey of
edge-on galaxies with relatively normal or disk-wide star forming
activity (T$\rm\ddot{u}$llmann et al.~2006a,b). In addition, there
might be an energy input threshold to form a multi-phase halo (Rossa
\& Dettmar 2003; Dahlem et al. 2006). In general, little is yet
known about the physical state of the extraplanar hot gas (e.g., the
pressure and volume filling factor) and its relationship with other
ISM components in the host halo.

The edge-on star-forming galaxy NGC 5775 (see Table \ref{basicpara}
for basic parameters) is well-suited for a detailed study of the
disk/halo interaction. Unlike a nuclear starburst galaxy, the star
formation in NGC 5775 is spread out across the galactic disk. This
allows us to explore the general disk/halo interaction without it
being overwhelmed by a powerful galactic nuclear superwind. Indeed,
this galaxy is one of the rare cases in which extraplanar emission
has been detected from essentially all diffuse interstellar
components (e.g., Lee et al. 2001). A number of extraplanar features
are resolved and found to exhibit coherent structures in \ion{H}{1},
radio continuum and H$\alpha$. NGC 5775 is interacting with its
companion NGC 5774, as they are bridged by \ion{H}{1} gas (Irwin
1994). In addition, the kinematics of the extraplanar diffuse
ionized gas has been carefully studied, showing a  vertical gradient
in azimuthal velocity, which is steeper than what is predicted by a
ballistic model of a disk/halo flow (Heald et al. 2006; Fraternali
et al. 2007). Hydrodynamic effects may thus be important (e.g.,
Barnab$\grave{e}$ et al. 2006), which may result from the
interaction of the flow with pre-existing halo gas. Previous X-ray
studies based on {\sl ROSAT} and {\sl XMM-Newton} observations of
NGC~5775 (Lee et al. 2001; T$\rm\ddot{u}$llmann et al.~2006a) have
revealed the presence of apparently diffuse X-ray emission which is
generally spatially related to the extraplanar features observed at
other wavelengths. However, these X-ray data of limited spatial
resolution do not allow for a clean detection and removal of
discrete sources or for a detailed morphological characterization of
the extraplanar hot gas and its connection to other components of
this galaxy.

In this paper, we present a {\sl Chandra} study of NGC~5775 and
confront the results with multi-wavelength data obtained from {\sl
Spitzer} and {\sl HST} as well as various ground-based observations.
These data allow us to examine various spatially resolved properties
of the disk/halo interaction. We describe
the observations and data reduction in \S~\ref{sec:datareduction} and
present our analysis and results in \S~\ref{sec:an_re}. We discuss
the implications of our results in \S~\ref{sec:discussion} and summarize
our findings in
\S~\ref{sec:summary}. Throughout this work errors are quoted at the
90\% confidence level.

\section{Observations and Data Reduction}{\label{sec:datareduction}}

The archival {\sl Chandra} observation of NGC 5775 (Obs. ID. 2940)
was taken on April 5, 2002, with a total effective exposure of 58.2
ks. In this work, we mainly utilize data from the ACIS-S3 chip,
although part of the data from the S2 chip is also used in the
source detection and imaging. We reprocessed the data using CIAO
v3.3 (the Chandra Interactive Analysis of Observations) following
the {\sl Chandra} ACIS data analysis guide. The light curve of the
observation indicates a quiescent background rate with no
significant flares. We created count and exposure maps in four bands
(0.3-0.7, 0.7-1.5, 1.5-3 and 3-7 keV) and use the ``stowed'' data
for the non-X-ray background subtraction after a count rate
normalization in the 10-12 keV range.

We performed source detection in the broad (B, 0.3-7 keV), soft (S,
0.3-1.5 keV) and hard (H, 1.5-7 keV) bands, following the procedure
detailed in Wang (2004). To study the diffuse X-ray emission, we
removed the detected discrete sources (\S~\ref{subsec:ps}) from the
maps. To do so, we excluded circular regions containing twice the
90\% enclosed energy radius (EER) around each source of a count rate
$(CR)\lesssim 0.01 {\rm~cts~s^{-1}}$. For brighter sources, the
source removal radius is further multiplied by a factor of $1+{\rm
log}(CR/0.01)$. Generally about 96-97\% of the source counts are
excluded in such a procedure.

Noticeably, the sky position of NGC~5775 is on the outskirts of the
North Polar Spur (NPS), a Galactic soft X-ray feature (Snowden et
al.~1995) that introduces an enhancement to the sky background.
Hence, in spectral analysis we adopted a ``double-subtraction''
procedure, similar to that used in Li, Wang \& Hameed (2007), to
determine the sky background.  A source-removed local background
spectrum is extracted from a region as shown in the left panel of
Fig. \ref{pointsrc}. The corresponding spectrum of the non-X-ray
contribution is extracted from the stowed data and subtracted from
the total background spectrum. We then characterize the remaining
sky signals by a combination of three characteristic components: a
thermal plasma model (APEC in XSPEC, with a temperature of $\sim$
0.1 keV) primarily for the soft X-ray emission of the Galactic halo
and the Local Hot Bubble, a second APEC (temperature $\sim$0.3 keV)
for the NPS emission, and a power-law (photon index of 1.4)
accounting for the extragalactic background. The sky background is
well fitted by these models over the 0.3-7 keV energy range. We
further verify the modeling by confirming that the amplitudes of the
three components are consistent with independent measurements (e.g.,
Willingale et al.~2003; Moretti et al.~2003). This background model
is added to the source models when fitting the non-X-ray
background-subtracted source spectra (\S~\ref{subsec:diffuse}), with
proper scaling of the sky area.

To assist the interpretation of the X-ray data, we incorporated
various relevant data sets for comparison. We obtained the {\sl
Spitzer} IRAC images of NGC~5775 from the {\sl Spitzer} archive. To
determine the net 8 ${\rm \mu}$m emission from dust, we divided the
background-subtracted, aperture-corrected (Reach et al.~2005) 4.5
${\rm \mu}$m image by a factor of 2.72 (Hunter, Elmegreen \& Martin
2006) and then subtracted it from the 8 ${\rm \mu}$m image to
account for the stellar contribution (the galaxy is unfortunately
absent in the 3.6 ${\rm \mu}$m channel in which the determination of
stellar contribution would be optimal).

The {\it HST} observations of NGC\,5775 were carried out on August
21 2005 with the Advanced Camera for Surveys (ACS) (GO\#10416; PI:
J.~Rossa) using the Wide Field Channel (WFC), which yields a
$202\arcsec \times 202\arcsec$ field of view. The narrowband F658N
filter observation was acquired in three {\it HST} orbits, while the
broadband F625W image was acquired in one {\it HST} orbit. More
observational details are given in Rossa et al. (2008). The {\it
HST}/ACS data were calibrated using the {\it CALACS} and {\it
Multidrizzle} packages. The former removes the various instrumental
signatures, including bias correction and flat-fielding. The latter
is used to correct for the geometric distortion of the ACS camera
and also performs cosmic ray rejection, and finally combines the
individual images, obtained in the specific dither pattern.

We also make use of \ion{H}{1} data of Irwin (1994), H$\alpha$ data
of Collins et al. (2000) and radio continuum data from Duric et al.
(1998) and Lee et al. (2001).

\section{Analysis and Results} {\label{sec:an_re}}

\begin{figure}[!h]
\centerline{ \epsfig{figure=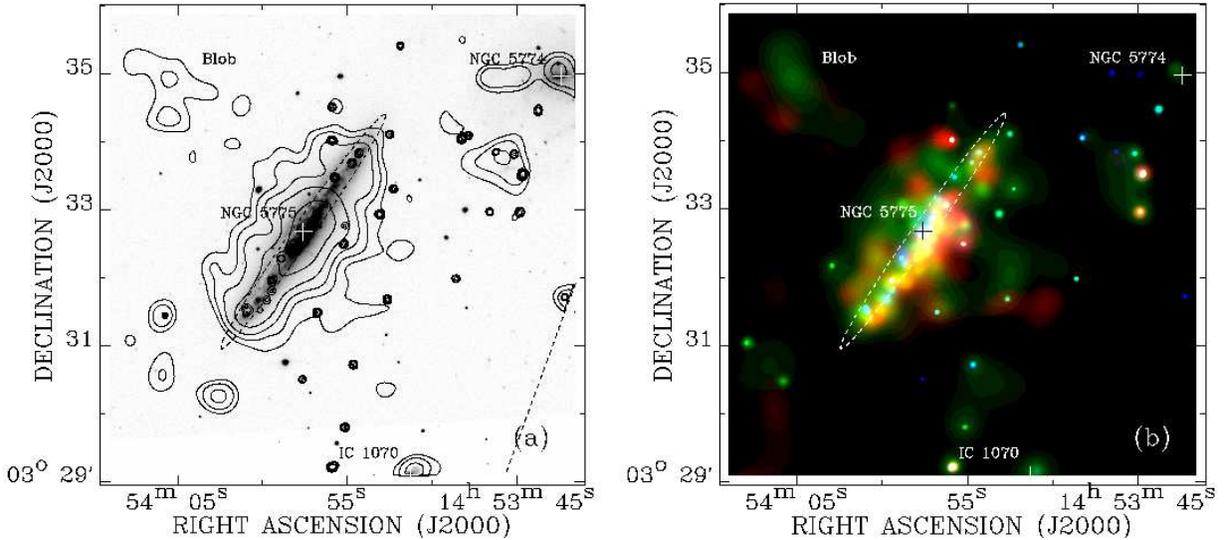,width=1.0\textwidth,angle=0,
clip=} } \caption{(a): ACIS-S 0.3-7 keV intensity contours overlaid
on the SDSS R-band image of NGC 5775 and its vicinity. The X-ray
intensity is adaptively smoothed with the CIAO {\sl csmooth} routine
with a signal-to-noise ratio of 3. The contours are at (2.5, 3.5, 5,
8, 15, 100 $and$ 500)${\times}10^{-3}{\rm~cts~s^{-1}~arcmin^{-2}}$.
The dashed line shows the edge of ACIS-S3 chip. The dashed ellipse
represents the $I_B = 25 {\rm~mag~arcsec^{-2}}$ isophote of the
galaxy. The plus signs mark the optical centers of NGC~5775 and its
companion galaxies NGC~5774 and IC~1070. (b): {\sl Chandra} ACIS-S
intensity image of NGC~5775 in tri-colors: red (0.3-0.7 keV), green
(0.7-1.5 keV), and blue (1.5-7 keV). Both 0.3-0.7 keV and 0.7-1.5
keV sub-bands trace primarily diffuse hot gas. The relative
strengths of the emission in the two sub-bands may be used to
characterize the temperature of the hot gas. The X-ray intensity is
adaptively smoothed with the {\sl csmooth} to achieve a
signal-to-noise ratio of 3 in all three bands.}\label{fig:optical}
\end{figure}

Fig. \ref{fig:optical}a shows ACIS-S 0.3-7 keV intensity contours of
NGC 5775 and its vicinity on an optical image, while Fig.
\ref{fig:optical}b shows a smoothed 3-color X-ray image. An
asymmetry of the diffuse emission is apparent in both the color and
brightness relative to the major axis close to the disk. This
asymmetry is clearly caused by the tilt of the galactic disk with
its nearer side toward the northeast. Aside from the X-ray emission
from NGC~5775 itself, apparently weak X-ray emission features also
appear to be associated with IC 1070 and NGC 5774. However, the
counting statistics of these detections are too low to allow for any
quantitative characterization. In particular, NGC 5774 is located in
the S2 chip that has a relatively lower sensitivity to the detection
of soft X-rays. The appearance of the soft X-ray emission from this
galaxy may be strongly perturbed by the tidal interaction with NGC
5775 (Lee et al. 2001). In the following we will concentrate on the
X-ray emission associated with NGC 5775, which in general represents
the combined contributions from discrete sources and truly diffuse
hot gas.

\begin{figure}[!h]
\centerline{ \epsfig{figure=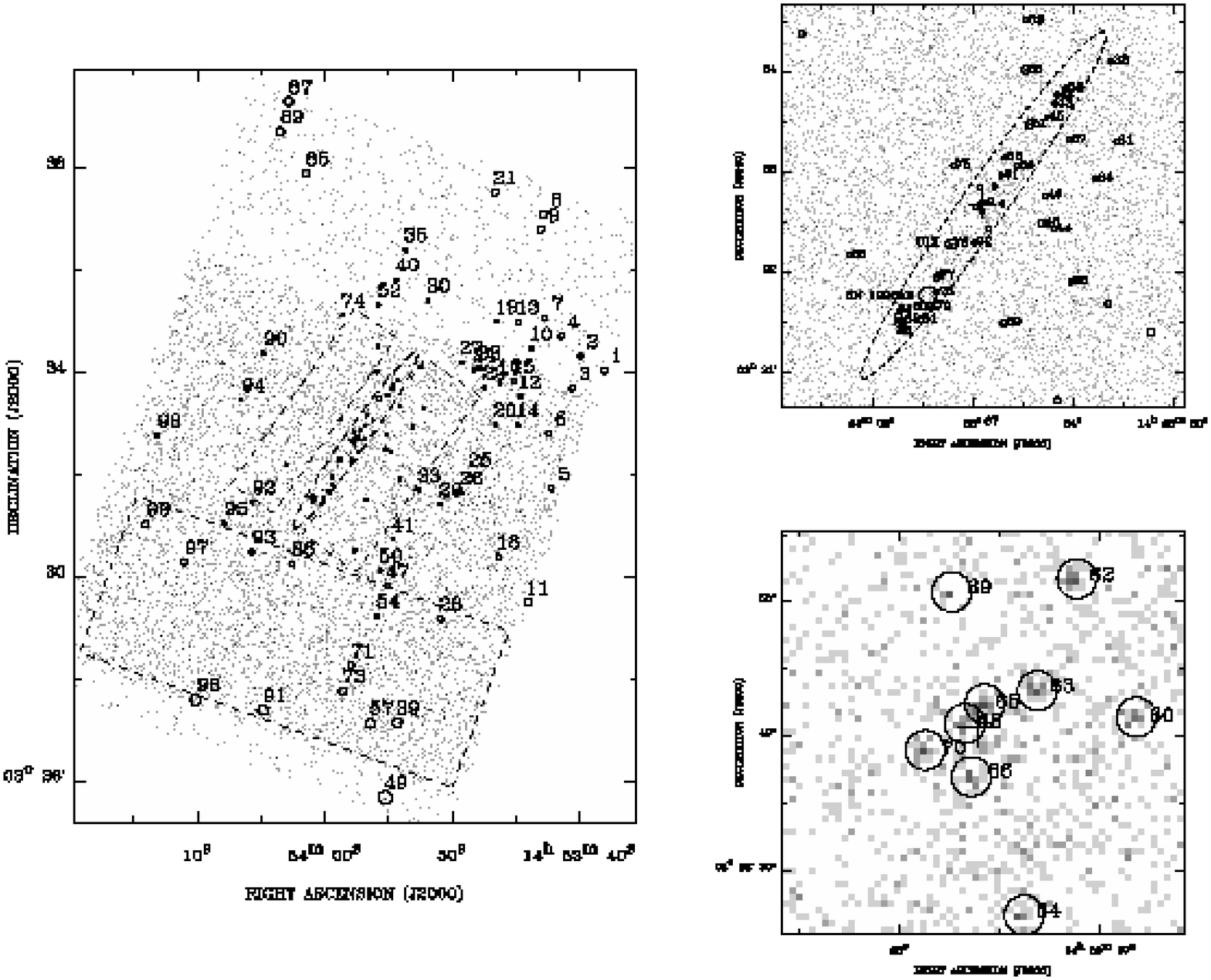,width=1.0\textwidth,angle=0,
clip=} } \caption{The \chandra\ images of NGC 5775 in the 0.3-7 keV
band. Detected sources are enclosed with circles of about 96\% PSF
EER and labeled with the numbers as in Table \ref{detpointsrc}. The
plus sign marks the center of the galaxy. {\sl Left panel}: The
dashed ellipse outlines the galactic disk region
(4\farcm2$\times$0\farcm44), while the boxes outline the halo and
background regions for spectral analysis (see text for details).
{\sl Top Right}: The central $4^\prime\times4^\prime$ field. In
particular, Source 76 is marked with a label ``ULX'', and the
position of SN 1996AE with a thick circle. {\sl  Bottom Right}:
Close-up of the central $0\farcm5\times0\farcm5$
field.}\label{pointsrc}
\end{figure}

\subsection{Discrete X-ray sources} {\label{subsec:ps}}

\begin{figure}[!h]
\centerline{ \epsfig{figure=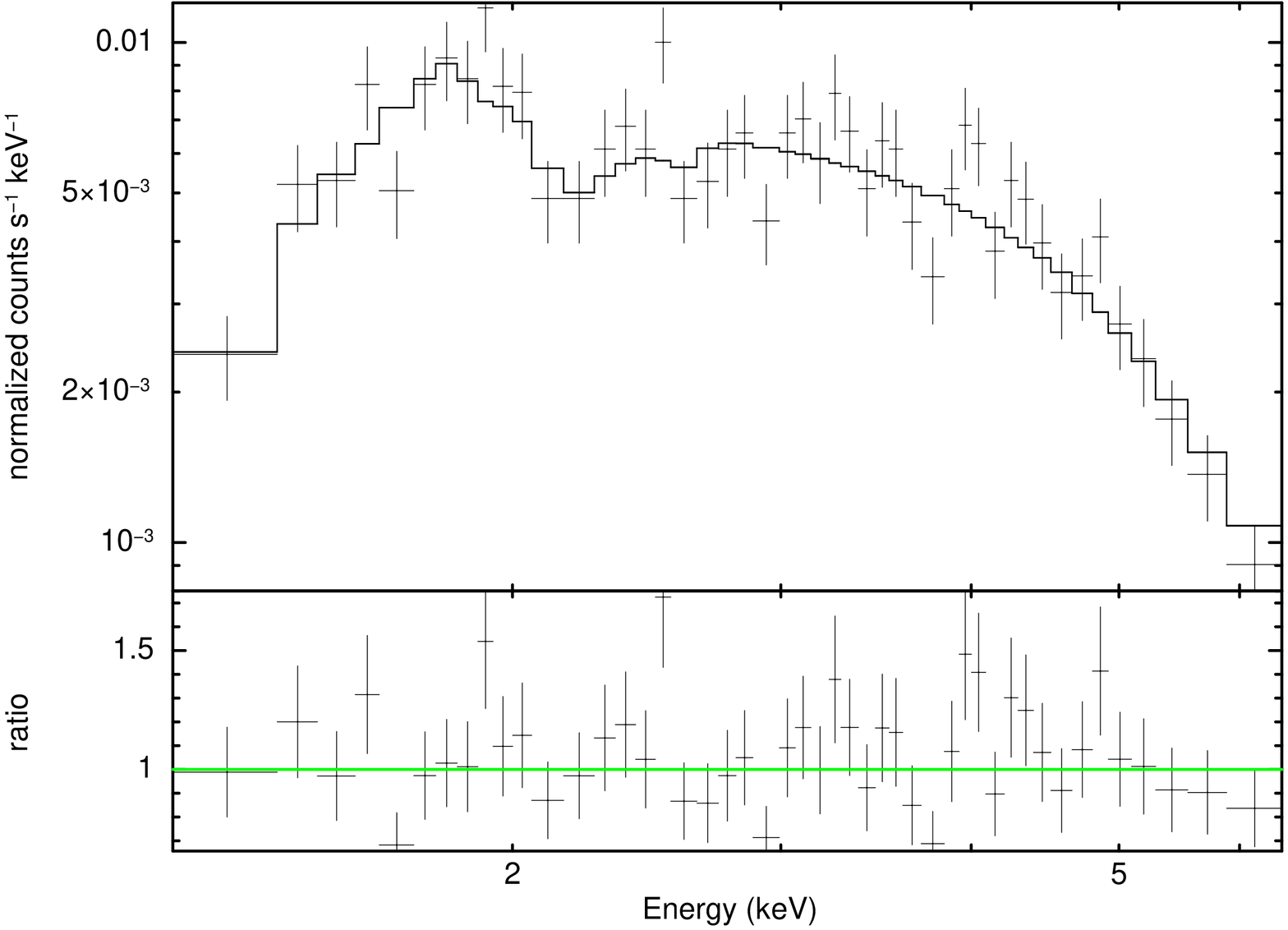,width=0.5\textwidth,angle=-90,
clip=} } \caption{The spectrum of the ULX (Source 76) fitted with an
absorbed power-law model (Table \ref{specfitpara}). The lower panel
shows the data-to-model ratio.}\label{ULXspectra}
\end{figure}

We detect 99 discrete sources in the field of NGC 5775
(Fig.~\ref{pointsrc}), 27 of which are within $D_{25}$ of the
galaxy, with a local false detection possibility of $\leq 10^{-6}$.
Table \ref{detpointsrc} summarizes information about these sources.
Among them, two sources have count rates greater than 20 cts
ks$^{-1}$: Sources 12 and 76. Source 12 is located at the outskirts
of the face-on spiral galaxy NGC 5774, showing a spectrum that can
be fitted by a power-law model (photon index of $\sim$2) with
absorption similar as the foreground value, hence it is likely an
X-ray binary in NGC 5774. Source 76 is located within the projected
galactic disk of NGC~5775, hence it is probably associated with this
galaxy. The spectrum of Source 76 can be well fitted by a power-law
with a photon index of $1.82_{-0.22}^{+0.31}$ and an absorption
column density of $2.86_{-0.37}^{+0.50}\times10^{22}\rm~cm^{-2}$
(Fig. \ref{ULXspectra}; Table \ref{specfitpara}). \ion{H}{1}
observations (Irwin 1994) indicate that the beam-averaged neutral
hydrogen column density is no higher than $10^{22}{\rm~cm^{-2}}$
within the disk region but substructure within the beam could result
in locally higher column densities.  Also, molecular gas in
star-forming regions may contribute additional absorption to Source
76. The fitted spectral model gives an absorbed (intrinsic) 0.3-7
keV luminosity of $2.4~(6.5)\times10^{40}\rm~ergs~s^{-1}$, putting
Source 76 in the class of ultra-luminous X-ray sources (ULXs).

The \chandra\ data show no significant X-ray emission from either
the nucleus of NGC~5775 or the remnant of SN 1996AE (Fig.
\ref{pointsrc}). An X-ray detection of the remnant was suggested by
T$\rm\ddot{u}$llmann et al. (2006a) based on an \xmm observation. We
speculate that this \xmm detection corresponds to our Source 76,
which is about 0\farcm5 from the remnant. In addition, the \xmm
source near the northwestern end of the galactic disk
(R.A.=$14^h53^m56^s$ and Dec.=$03^\circ34^\prime00\farcs0$;
T$\rm\ddot{u}$llmann et al.~2006a, Fig.~10 therein) is most likely
our Source 55 --- the fourth brightest source in our detection list.
This source has an R-band counterpart with a magnitude of
$\sim21\rm~mag$. Its X-ray spectrum is well fitted by a power law
with a photon index of $1.2\pm0.3$.

We have further analyzed the accumulated spectrum of detected sources
in the disk region (Fig.~\ref{pointsrc}) and these can be well fitted
with an absorbed power-law model with a photon index of $\sim$1.3.

\subsection{Diffuse X-ray emission} {\label{subsec:diffuse}}

\begin{figure}[!htb]
\centerline{
  \epsfig{figure=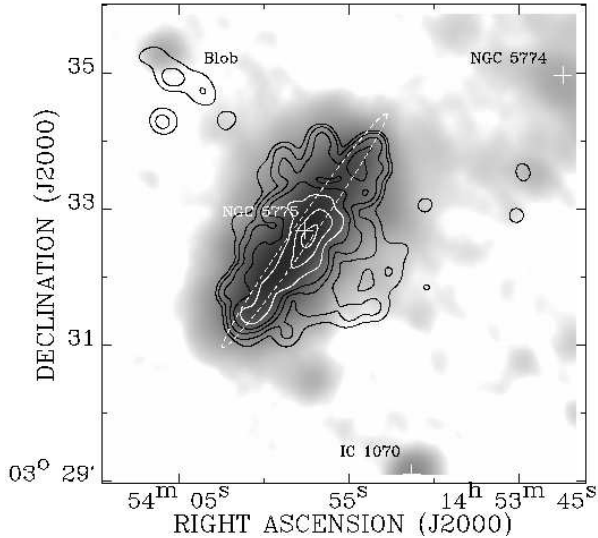,width=0.5\textwidth,angle=0}
} \caption{{\sl Chandra} ACIS-S 0.3-1.5 keV intensity contours of
the diffuse emission overlaid on the 20 cm radio continuum map of
NGC~5775 (Lee et al.~2001). The contours are at (2.0, 2.5, 3.5, 5,
8, 15 and 25) ${\times}10^{-3}{\rm~cts~s^{-1}~arcmin^{-2}}$.
\label{fig:pointsrcblob}}
\end{figure}

Fig. \ref{fig:pointsrcblob} shows an image of the discrete
source-removed ``diffuse'' soft X-ray emission, compared with the 20
cm radio continuum map of NGC 5775. In addition to the presence of
the diffuse emission clearly associated with NGC 5775, an isolated
``blob'' of X-ray emission also appears northeast of the galaxy. The
far side of this blob reaches a projected distance of $\sim$3\farcm5
($\sim$25 kpc) from the galactic center. As evident in Fig.
\ref{fig:optical}b, this feature is only apparent in the soft
(0.3-1.5 keV) band. Although the detection is formally at a statistical
confidence of $\sim9.4\sigma$, the feature has a low surface
brightness. We cannot completely rule out the possibility that it
may be an artifact, caused by the systematic uncertainty in the
background subtraction, flat-fielding, and/or faint source confusion.
There is no evidence for any group or cluster of galaxies in this
field. Therefore, the feature, if real, is likely
to be associated with NGC~5775. In the opposite side, there is a
``shell-like'' feature extending to a projected distance of
$\sim$1\farcm5 ($\sim$11 kpc) from the galactic plane.

\begin{figure}[!htb]
\centerline{
  \epsfig{figure=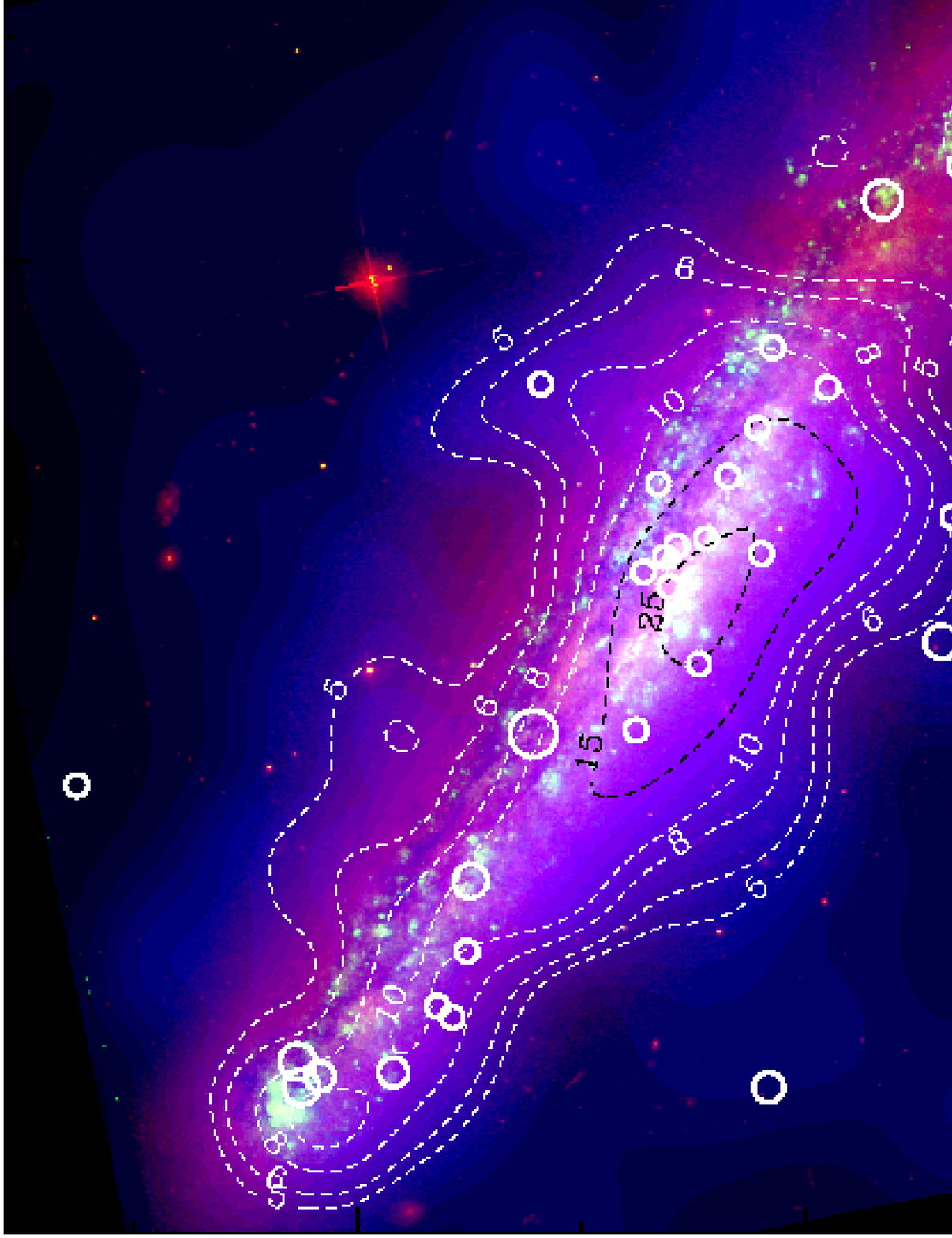,width=0.7\textwidth,angle=0}
} \caption{Diffuse X-ray intensity and the {\sl HST/ACS} images:
R-band (red), H$\alpha$ (green), and 0.3-1.5 keV (blue). The
smoothing of the X-ray image is the same as in
Fig.~\ref{fig:pointsrcblob}. The contour levels are marked on the
image in units of $10^{-3}{\rm~cts~s^{-1}~arcmin^{-2}}$. The white
circles mark the discrete sources detected in the same X-ray band as
shown in Fig. \ref{pointsrc}. \label{fig:x_hst.ps}}
\end{figure}

\begin{figure}[!htb]
  \centerline{
      \epsfig{figure=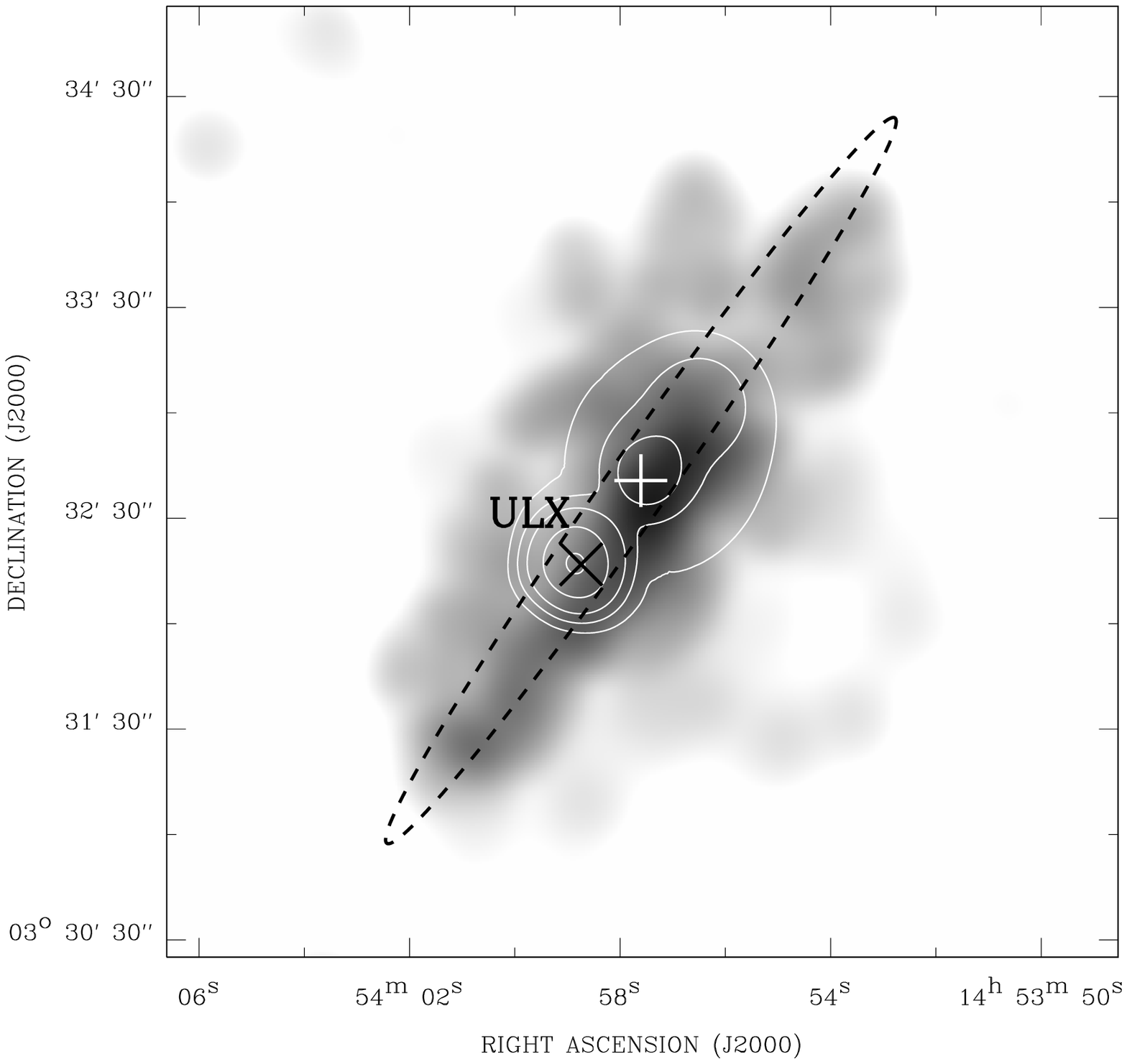,width=0.5\textwidth,angle=0}
  }
 \caption{Comparison of {\sl Chandra} ACIS-S diffuse intensity
images of NGC~5775 in the 1.5-7 keV (contours) and 0.3-1.5 keV (gray-scale)
bands.
The 1.5-7 keV intensity is adaptively smoothed to achieve a signal-to-noise ratio higher than 3, while the 0.3-1.5 keV intensity is the same as in
Fig.~\ref{fig:pointsrcblob}. The
contours are at (0.5, 1, 2, 5, 10)
${\times}10^{-3}{\rm~cts~s^{-1}~arcmin^{-2}}$. The position of the ULX
is marked with a cross to illustrate the presence of its residual contamination.} \label{fig:Xh_multi}
\end{figure}

Fig. \ref{fig:x_hst.ps} presents the diffuse 0.3-1.5 keV intensity
map of NGC~5775, together with {\sl HST/ACS} images. Current star
formation, as traced by the H$\alpha$ emission, seems to occur
throughout the galactic disk, but is clearly more active toward its
inner region and in a few giant complexes, in which a greater number
of discrete X-ray sources also reside.
Along the major axis, a dust lane, which is
seen as a dark band in
the R-band image, causes the soft X-ray absorption
and hence the asymmetry in the soft X-ray intensity distribution.
Several X-ray intensity contours have been
displayed to aid in illustrating the connection between
\ion{H}{2} complexes in the disk and the broader scale diffuse
X-ray emission.  The giant \ion{H}{2} complex near
the far NW end of the major axis is a good example.
%X-ray emission enhancements to the giant \ion{H}{2} complexes.
Diffuse emission in the 1.5-7 keV band (Fig.~\ref{fig:Xh_multi}) is
only present in the central region ($ r \lesssim 5$ kpc) along the
galactic disk. The enhancement at the position of the ULX is mostly
due to its residual contamination (the PSF wing outside the
source-removal circle).

\begin{figure}[!htb]
\centerline{
  \epsfig{figure=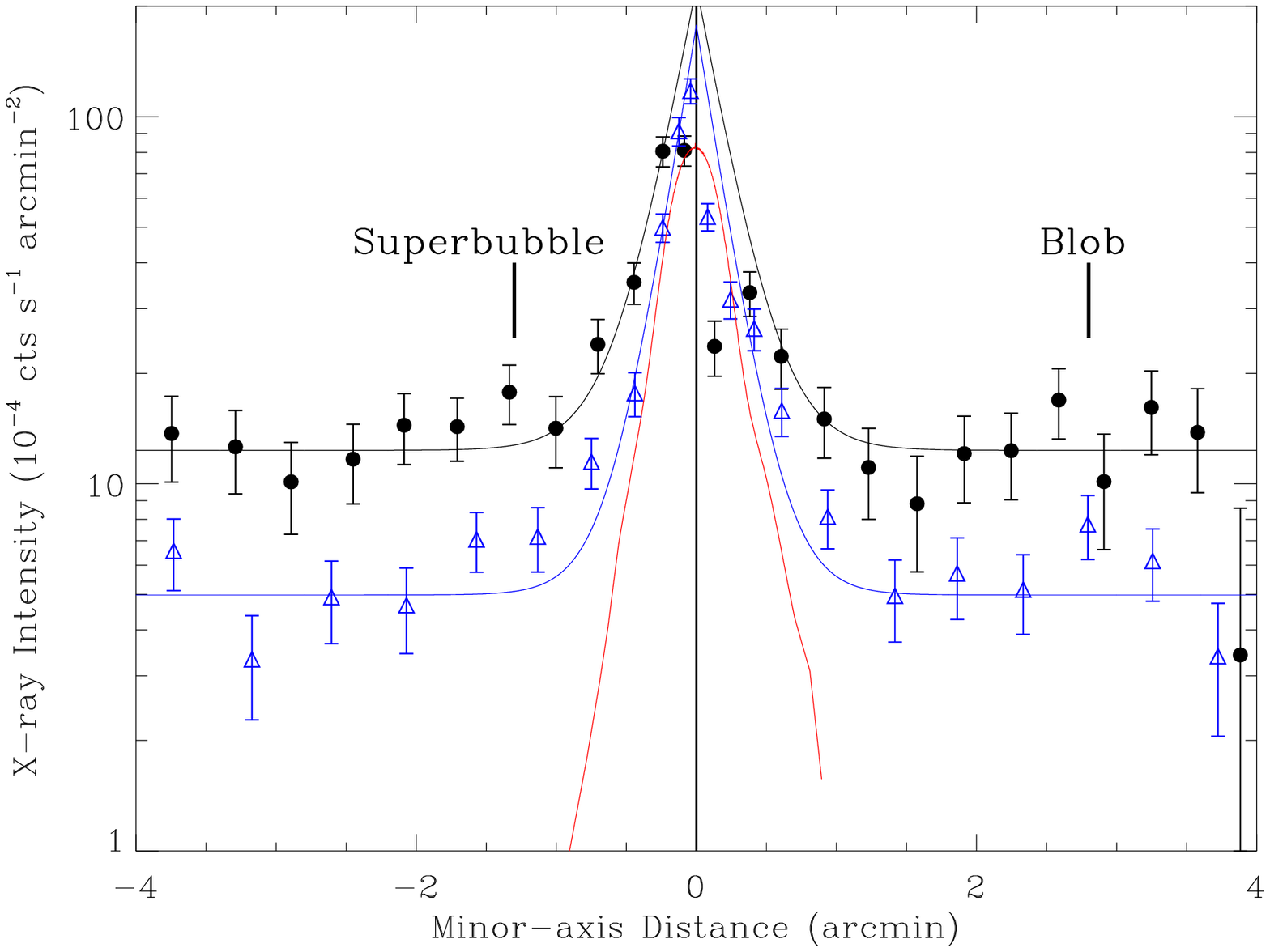,width=0.7\textwidth,angle=0}
} \caption{{\sl Chandra} ACIS-S intensity distributions of the
diffuse emission as a function of the distance off the major axis of
NGC~5775. The data are extracted in the 0.3-0.7 keV ({\sl black
filled circles}) and 0.7-1.5 keV ({\sl blue triangles}) bands
(southwest as negative). The full width along the major axis of the
disk used for averaging the intensity is 200$^{\prime\prime}$.
Spatial binning achieves a counts-to-noise ratio greater than 10 and
a minimum step of 5$^{\prime\prime}$. The red solid curve shows the
corresponding \ion{H}{1} intensity with arbitrary normalization. The
central vertical line represents the position of the major axis. The
black and blue curves on the negative side show the best-fit
exponential law plus a local background, while on the positive side,
the fitted curves have the same parameters and are only plotted for
comparison.} \label{fig:profile_major}
\end{figure}

Fig.~\ref{fig:profile_major} shows the vertical diffuse X-ray
intensity profiles as a function of the distance from the galaxy's
major axis. The profiles are calculated separately for the 0.3-0.7
and 0.7-1.5 keV bands and are averaged within a slice that is
200$^{\prime\prime}$ wide along the major axis. Within an off-disk
distance of $\sim$0\farcm5 to the northeast (positive side in
Fig.~\ref{fig:profile_major}), a dip is apparent in the intensity
profiles: deeper in the 0.3-0.7 keV band than in the 0.7-1.5 keV
band and is fully consistent with the X-ray absorption by the tilted
disk (Irwin 1994). We fit the profiles with an exponential law:
$I(z)=I_ge^{-|z|/z_0}$, where $I_g$ is the intensity at the major
axis and $z_0$ is the vertical scale height. A local sky background
is adopted to be
$12.3\pm0.7~(5.0\pm0.3)\times10^{-4}\rm~cts~s^{-1}~arcmin^{-2}$ in
the 0.3-0.7 (0.7-1.5) keV band, estimated from a background region
as described in \S \ref{sec:datareduction}. Since on the positive
side, the profile is seriously affected by the absorption with a
peak column density $N_{\rm{HI}} \sim 10^{22} {\rm~cm^{-2}}$ in the
galactic plane, we only fit the profile in the negative side with
$z\lesssim-0\farcm2$ to minimize the residual absorption effect of
the disk. The 0.3-0.7 (0.7-1.5) keV scale heights are
$\sim0\farcm21$ ($0\farcm18$) in the southwestern side. Accounting
for the absorption effect, the intrinsic diffuse soft X-ray emission
appears to be fairly symmetric relative to the major axis of the
galaxy. Nevertheless, imprints of large-scale extraplanar features
are present; the most noticeable are the southwest shell-like
feature and the northeast blob (Fig. \ref{fig:optical}, Fig.
\ref{fig:pointsrcblob}; see \S\ref{sec:discussion} below). These
features produce local enhancements, although their significance is
already diluted in Fig.~\ref{fig:profile_major} because of the
average over the width of the broad slice.

\begin{figure}[!htb]
\begin{center}
\centerline{
    \epsfig{figure=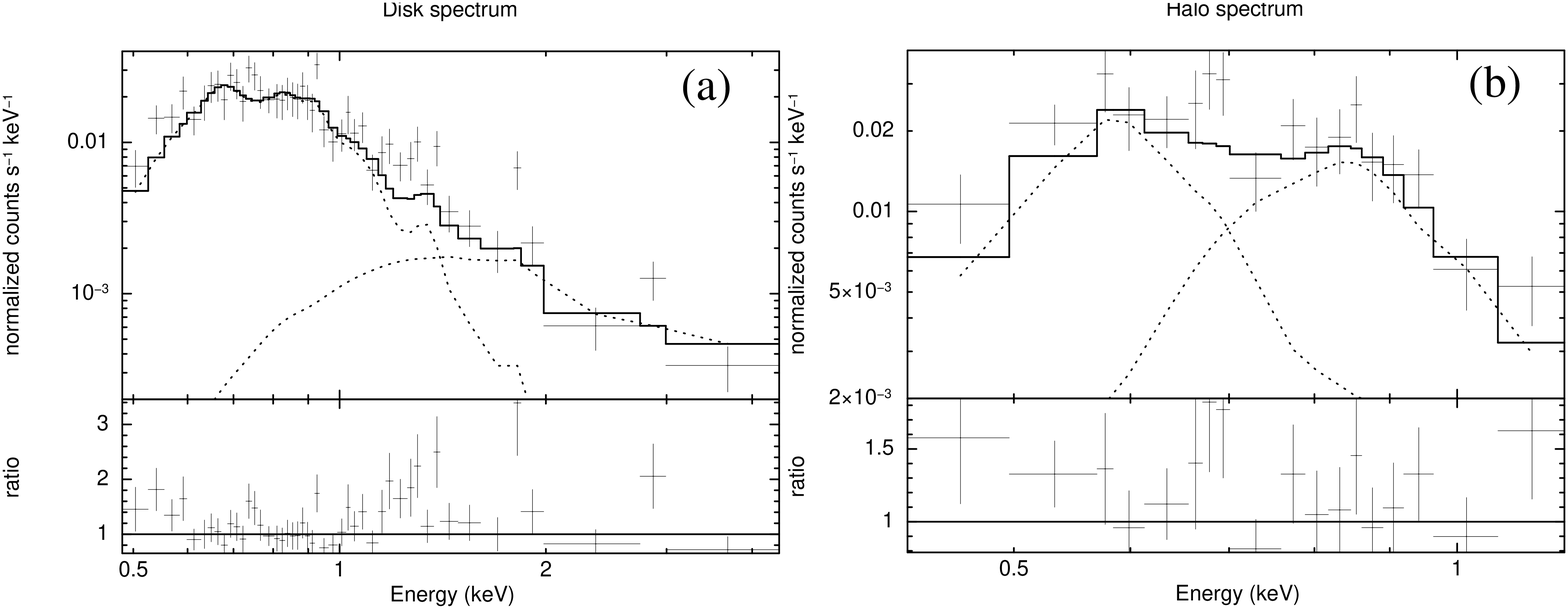,width=1.0\textwidth,angle=0, clip=}
} \caption{Diffuse X-ray spectra of the (a) disk and (b) halo
regions after non-X-ray and sky background subtraction. The spectra
are binned to achieve a signal-to-noise ratio better than 3 with
respect to the best-fitted local background. The fit parameters are
summarized in Table \ref{specfitpara}. The solid and dotted curves
represent the best-fit models and their components, respectively.
See text for details.}\label{spectra}
\end{center}
\end{figure}

We extract spectra from the disk and halo regions in NGC 5775 for
the diffuse X-ray emission (Fig.~\ref{pointsrc}). The X-ray emission
from regions northeast to the major axis of the galaxy is subject to
the absorption by the tilted cool gas disk. The \ion{H}{1} disk, in
particular, is likely to be thicker at large galactocentric radii,
responsible for the projected broad vertical \ion{H}{1} distribution
seen in Fig. \ref{fig:X_multi}d. The near side of the \ion{H}{1}
disk can thus effectively absorb X-ray from the extraplanar X-ray
emission from the northeast halo of the galaxy, but should be tilted
largely away from sight-lines toward the southwest part of the halo.
The \ion{H}{1} projected in this latter part of the sky is located
on the far side of the disk and should produce little absorption on
the extraplanar X-ray emission. Therefore, we collect the X-ray
spectral data only from southwest halo and large height of northeast
halo, as outlined in Fig.~2. The spectral analysis below (e.g., Fig.
8b) and the spatial distribution of the extraplanar X-ray emission
(e.g., Fig. 7) also show no indication for any absorption, in
addition to that caused by the foreground Galactic column density.

We fit the disk and halo spectra with a thermal plasma model (XSPEC
MEKAL, with the abundance fixed at solar). At energies above 1.5 keV
of the disk spectrum, the residual counts spilled outside the
source-removal regions (especially the ULX) and the unresolved point
sources below our detection limit dominate, so we add a power law
(PL) component to account for this effect (a single MEKAL model gives
very poor fit with reduced  $\chi^2\sim3.1$). Intrinsic absorption is
considered in the fit to the disk spectrum and the fitted value is
consistent with the \ion{H}{1} data (Fig. \ref{fig:profile_major}).
Due to the limited statistics of high energy counts, the photon
index of the PL component cannot be well constrained and is thus
fixed to the value from the fit to the accumulated source spectrum
(\S~\ref{subsec:ps}). The fitted temperature of the diffuse plasma
is $\sim$0.2 keV in the disk. The thermal (PL) component of the disk
spectrum gives an intrinsic 0.3-2 keV luminosity of $\sim66~(1.4)
\times10^{39}\rm~ergs~s^{-1}$. This model fit provides only
a simple characterization of the spectral shape of the disk X-ray emission.
The fitted individual model parameters probably do not have much
physical meaning. For example, the power law component may contain a
significant contribution from hotter diffuse gas that may be expected
in the galactic disk. But the limited counting statistics of the data
and the edge-on orientation of the galaxy make it impossible to
significantly improve the analysis.

Since the stellar contribution in the halo is much lower than that
in the disk, we do not include a PL component for the halo spectrum.
But a single MEKAL model gives relatively poor fit with reduced
$\chi^2\sim1.7$, so we use a two-temperature model for the halo,
again with abundances fixed at solar. The absorption is fixed at the
foreground value. The 0.3-2 keV intrinsic luminosity of the low
(high) temperature component is $\sim2.1~ (1.4)
\times10^{39}\rm~ergs~s^{-1}$. The fitted spectra are shown in
Fig.~\ref{spectra}, while the fitted parameters are summarized in
Table \ref{specfitpara}.

\section{Discussion}\label{sec:discussion}

\subsection{Connection of extraplanar hot gas to disk star formation}\label{sec:XFeature}

\begin{figure*}[!htb]
\centerline{
  \epsfig{figure=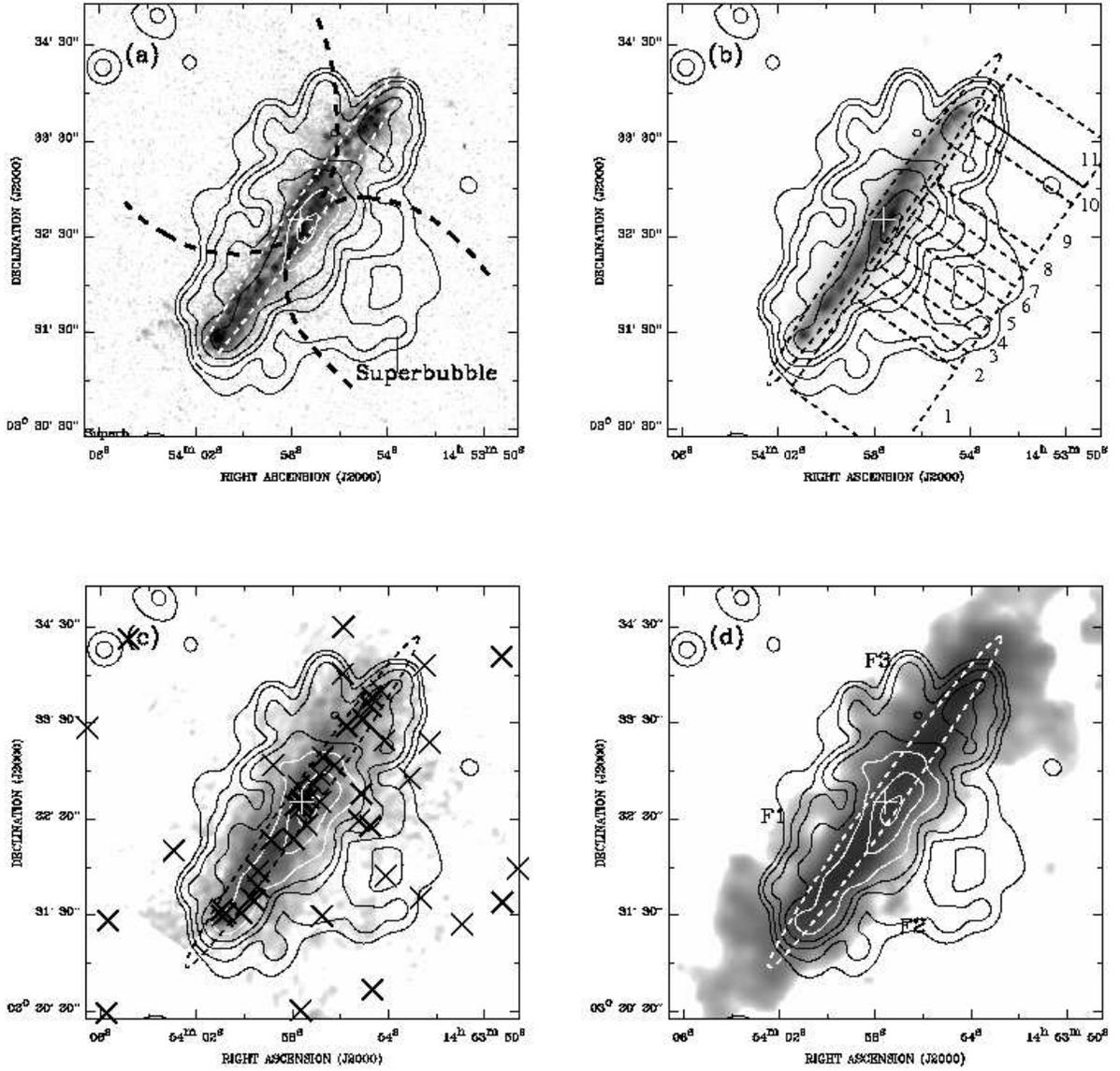,width=1.0\textwidth,angle=0} }
\caption{{\sl Chandra} ACIS-S 0.3-1.5 keV intensity contours of the
diffuse emission overlaid on the H$\alpha$ (a; Collins et al.~2000),
{\sl Spitzer} 8 ${\rm \mu}$m(b), 20 cm radio continuum (c; Lee et
al.~2001) and \ion{H}{1} column density (d; Irwin 1994) of NGC~5775.
The X-ray intensity is the same as in Fig.~\ref{fig:pointsrcblob}.
The contours are at (2.0, 2.5, 3.5, 5, 8, 15, 25)
${\times}10^{-3}{\rm~cts~s^{-1}~arcmin^{-2}}$. The ``X-shape''
feature discussed in \S\ref{sec:XFeature} is illustrated in (a); the
extraplanar regions used to calculated soft X-ray intensities in
Fig. \ref{fig:X_IR} are plotted in (b) with the region numbers marked
on the side; the positions of the detected sources are marked with
{\sl crosses} in (c); and the F features of Lee et al. (2001) are marked
in (d).} \label{fig:X_multi}
\end{figure*}

In Fig.~\ref{fig:X_multi} (also see
Figs.~\ref{fig:pointsrcblob} and \ref{fig:x_hst.ps}), we show the
comparative morphology of the soft X-ray emission with images in
other wavebands. The in-disk soft X-ray emission generally shows spatial
similarities with star formation regions as traced by H$\alpha$, IR
and/or radio emission. It is not surprising to have such
similarities, since diffuse hot gas and unresolved young stellar
objects, both closely related to star-forming regions, contribute to
the soft X-ray emission. The extraplanar X-ray emission arises from
truly diffuse hot gas. Thus, in general, we expect a spatial
relationship between the extraplanar hot gas and underlying
star-forming regions if the extraplanar gas is supplied and
regulated by underlying star-forming processes. Differences in
spatial resolution for the plots of Fig.~\ref{fig:X_multi} make
specific connections to underlying star forming regions
 more difficult to trace.  An examination of
Fig.~\ref{fig:x_hst.ps}, however, indeed does show some correlation.
For example, as indicated in Sect.~\ref{subsec:diffuse}, the in-disk
NW giant \ion{H}{2} complex appears connected to the extraplanar X-ray
emission and the central X-ray contours are wider around the central
star forming regions of the galaxy. The giant \ion{H}{2} in-disk
complex on the far SE end of the disk does not appear to have an
extra-planar X-ray counterpart in this figure, but the X-ray results
of T{\"u}llmann et al. (2006) reveal extraplanar X-ray emission
associated with a string of \ion{H}{2} regions off the plane on this
side as well.  Again, this argues for a connection between
extraplanar X-ray emission and in-disk star formation.

On larger scales,
several extraplanar features are resolved with the superb resolution
of {\sl Chandra} and other multi-wavelength observations (Fig.
\ref{fig:X_multi}). Here we discuss the three most prominent
plume-like features, originally identified
in \ion{H}{1}
observations (Irwin 1994; Lee et al. 2001; labelled F1, F2, and F3 in
Fig.~\ref{fig:X_multi}) and later found to have counterparts or extensions in
other wavebands (e.g. Lee et al. 2001; Brar et al. 2003).
F1 and F2,
being opposite to each other with respect to the disk, apparently
protrude from an intense star-forming region, within which the ULX
is located. On the other hand, in the disk region underneath F3, all
components of the ISM, except for \ion{H}{1}, show a local minimum,
indicating a low current star formation rate (Lee et al. 2001), possibly
as a result of a previous star forming episode which has produced an
in-disk disruption.
F3 seems to
be associated with an extraplanar H$\alpha$ filament, within which
several \ion{H}{2} regions reside (Lee et al.~2001).
Together with the disk-halo connections discussed above
(cf. Fig.~\ref{fig:x_hst.ps}), they collectively
indicate that local star forming regions away from the
nucleus affect halo characteristics
above and below the plane.

In addition, however, these features are located near the boundaries
of the so-called ``X-shape'' which designates conic structures
(marked in Fig. \ref{fig:X_multi}a) and was claimed to be present
from previous X-ray (T$\rm\ddot{u}$llmann et al. 2006a) and radio
(T$\rm\ddot{u}$llmann et al. 2000) observations. The shell-like
X-ray feature to the SW of the major axis also seems to be part of
the X-shape structure, representing an outflow from the galactic
center region to a vertical height of $\sim$1\farcm5 ($\sim$11 kpc)
(Fig. \ref{fig:X_multi}).  Thus, NGC~5775 appears to be similar to
the starburst galaxy, NGC~253 (e.g. Vogler \& Pietsch 1999; Pietsch
et al. 2001; Strickland et al. 2000) with its nuclear conical
outflow, yet it also has smaller-scale outflows associated with
specific in-disk regions.  As a result, it is not always
straightfoward to interpret the origin of specific extraplanar
features.  An example is the large northeastern blob shown in Fig.
\ref{fig:optical} and Fig. \ref{fig:pointsrcblob} which seems to
have a radio continuum counterpart. It is roughly above the F3
disturbance but could also be a result of the nuclear outflow. It is
located about twice as far away from the galactic disk than the
shell-like feature on the opposite side.

\subsection{The sub-galactic scale ``X-SF'' relation}\label{sec:X-SF}

\begin{figure}[!htb]
\centerline{
  \epsfig{figure=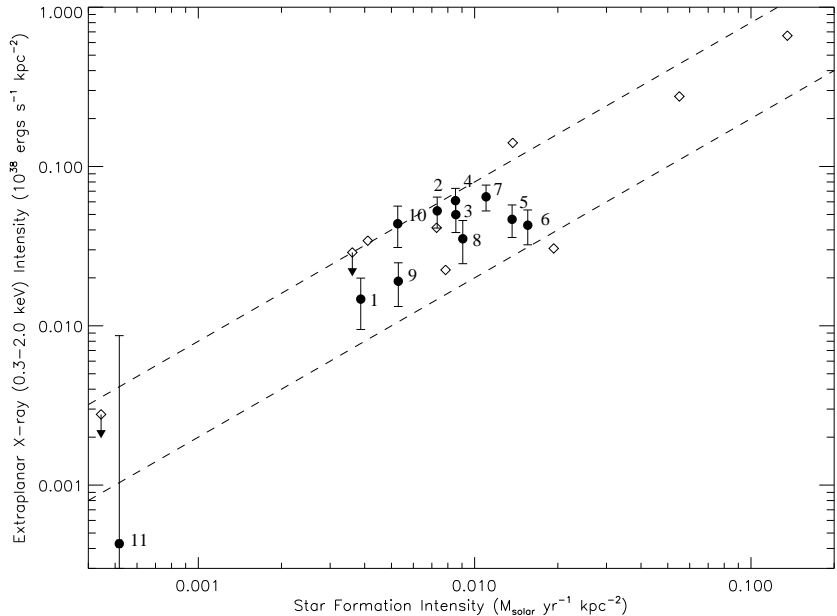,width=0.7\textwidth,angle=0}
} \caption{In-disk ($|z|\leq$~0\farcm2) star formation intensity
calculated from 8 $\rm \mu m$ intensity versus extraplanar
(-1\farcm5~$\leq z\leq$~-0\farcm2) soft X-ray intensity. The soft
X-ray intensity is obtained in 0.3-1.5 keV band and converted to
0.3-2 keV band for comparison with the Strickland et al. (2004a)
sample. The binning slices of the extraplanar regions are shown in
Fig. \ref{fig:X_multi}b, while the region numbers are marked on the side.
The star formation intensity is calculated just underneath
these slices with $|z|\leq$~0\farcm2. See text for details. Filled circles are
measurements from this work. Open diamond symbols denote
measurements from the Strickland et al. sample galaxies, in which
the X-ray luminosity of NGC 4244 and NGC 6503 are estimated with
upper limits and NGC 4945 is excluded due to its high foreground
absorption. Strickland et al.'s sample has a typical error in the y
axis of less than $10^{36}\rm~ergs~s^{-1}~kpc^{-2}$. Only NGC 1482
(the galaxy with the second highest extraplanar X-ray intensity in
their sample) has a larger error of about
$3\times10^{36}\rm~ergs~s^{-1}~kpc^{-2}$. The two dashed lines show
the approximate range of the data, which is
$I_{0.3-2.0{\rm~keV}}(10^{38}{\rm~ergs~s^{-1}~kpc^{-2}})\sim(2-8)~I_{SF}~({\rm
M_\odot~yr^{-1}~kpc^{-2}})$. } \label{fig:X_IR}
\end{figure}

The relation between the integrated extraplanar soft X-ray intensity
and the in-disk star formation intensity (hereafter referred to as
an ``X-SF'' relation) has been demonstrated for active star-forming
galaxies (Strickland et al.~2004b; T$\rm\ddot{u}$llmann et
al.~2006b). Fig. \ref{fig:X_IR} compares this relation with a
sub-galactic ($\sim\rm kpc$) scale version derived from our current
work.

Since H$\alpha$ emission may suffer from strong absorption in this
edge-on case, we estimate the star formation rate from the 8 $\rm
\mu m$ intensity using Wu et al. (2005)'s relation. We then convert
the X-ray intensity to intrinsic luminosity using the halo spectral
model (Table \ref{specfitpara}). Different in-disk regions should
have different line-of-sight depth, so when calculating in-disk area
we assume the disk has a circular shape with diameter of $D_{25}$.
To correct for projection, the inclination angle is adopted to be
$\sim86^\circ$ as listed in Table \ref{basicpara}. Note that,
although we use only the X-ray data in the southern halo to minimize
absorption effects, Fig. \ref{fig:X_IR} assumes the two sides have
the same intrinsic luminosity.

The consistency of the ``X-SF'' relations found inside NGC~5775 and
inferred from integrated galactic properties indicates that the
efficiency of converting stellar feedback to the extraplanar hot gas
does not change significantly from $\sim$kpc to galactic scales. The
threshold for the presence of the extraplanar X-ray emission on
these scales seems to be $\sim 4\times10^{-3}~\rm
M_{\odot~}yr^{-1}~kpc^{-2}$, consistent with previous findings based
on galaxy samples (Strickland et al. 2004b; T$\rm\ddot{u}$llmann et
al. 2006b). Fig.~\ref{fig:X_IR} also suggests that the conical
nuclear outflow is not strongly perturbing the extraplanar X-ray
emission that is above and below specific in-disk regions.

\subsection{Physical and dynamic states of the extraplanar hot gas}\label{sec:Hotgas_physical}

In \S\ref{subsec:diffuse}, we have shown that the halo spectrum
consists of two hot gas components. There are now at least three ISM
components co-existing in the halo of NGC 5775: the warm ionized gas
traced by H$\alpha$ emission, and the two hot gas components.
Inferred from H$\alpha$ and optical spectroscopic observation, the
warm ionized gas has a temperature of $\sim10^4~\rm K$ (Collins \&
Rand 2001) and density of $\sim0.1~\rm cm^{-3}$ (Collins et al.
2000). The warm gas likely has a low filling factor compared with
the hot gas if a thermal pressure balance is established between the warm and
hot gas. Strickland et al. (2002) presented several disk/halo
interaction models to interpret the multi-phase halo in NGC 253. In
their models, H$\alpha$-emitting warm ionized gas can be produced
through interaction between the outflow and the pre-existing gas in
the halo, from the cooling of swept up outer shells, or simply
by being dragged from the disk. Soft X-ray emission can arise from
 shock heated halo gas or from the conductive interfaces or
turbulent mixing layers between cold halo clouds and the tenuous,
low emissivity, and hot superwind.

Some parameters of the two hot gas components are listed in Table
\ref{halopara}. These parameters are inferred from the spectral
fitting parameters listed in Table \ref{specfitpara}, with the
assumption that the total filling factor of the hot gas is $\sim1$
and the two hot gas components reach the pressure balance with each
other. We expect that hot gas is produced mainly in the disk and
then flows into the halo. The temperature of the gas depends on both
the efficiencies of the mass loading as well as the radiative
cooling and other energy dissipation processes (e.g., via cosmic
rays and kinetic energy). Because the mass efficiency of star
formation is only $\sim 10\%$ and the remaining molecular gas is
dispersed on a time scale of $t \sim 10^7$ yrs, substantial
massloading to hot gas is expected in recent star forming regions.
With the presence of the dense gas, radiative and mechanical cooling
can also be important. In contrast, the heating of gas by
core-collapsed SNe can be more efficient in older star formation
regions ($10^7 \lesssim t \lesssim 5 \times 10^7$ yrs; the upper
limit corresponds to the lifetime of an 8$M_\odot$ star). Therefore,
gas in an ongoing or recent star forming region can be cooler ---
explaining the low-temperature component of the hot gas ---  than in
those older star formation regions, which are probably responsible
for the hotter component. The lower temperature hot gas can also be
produced in the halo at interfaces between the hot gas outflow and
the pre-existing cooler gas. Radiative cooling generally has so long
time scale for both hot gas components that it cannot be very
important in the halo (Table \ref{halopara}), at least in the
vicinity of the galactic disk.

The rim-brightened soft X-ray-emitting bubble likely represents the reverse-shocked
outflow in the interior of an outer cool shell (Strickland et al. 2002).
This cool gas shell could be formed in or near the disk, and then
expand into the halo driven by the high pressure interior.
During the expansion of this cool shell, the magnetic field may
play an important role in preventing the Rayleigh-Taylor instability
which would break the shell, given that the magnetic pressure may be
comparable with the thermal pressure of the halo hot gas (see
below).

The northeast blob may be compared with a similar feature ``ridge''
detected in M 82 (Lehnert, Heckman \& Weaver 1999).
The blob, elongated along the major
axis instead of perpendicular to the major axis, is about
twice as high as the M 82 ridge. Given that NGC 5775 does not have
as strong a superwind as detected in M 82, and there is no clear
sign of a cool gas counterpart for the blob, we speculate that it
may represent a bulk of outflowing hot gas --- an evolved outflowing
wind bubble --- generated in an early episode
of massive star formation in the nuclear region of the galaxy.
The relatively high concentration of the hot gas
at such a large off-disk distance may be due to the confinement of a
magnetic field loop, which would still be anchored in the disk. A
magnetic field must be present because of the observed radio
continuum counterpart. Data of higher quality is needed to truly
understand the nature of the blob.

Although a magneto-hydrodynamic model of the nuclear outflow is
beyond the scope of this paper, we can still make some general statements
about halo conditions.
 For example, the magnetic pressure
in the halo can be estimated from
${\rm P}_B\,\sim10^{-12}~[B/(5~\mu G)]^2~f_B^{-4/7}\rm~dyn~cm^{-2}$,
where $B$ is the magnetic field strength in units of
$\mu G$ and $f_B$
is the volume filling factor of the magnetic field.  In this relation,
the magnetic field strength is assumed to have been determined from
 multi-frequency
radio continuum maps, adopting an approximate energy equipartition between
relativistic particles and the magnetic field.
 Using
the radio continuum and spectral index maps of Duric et al. (1998)
and an appropriate geometric model (see Irwin et al. 1999 and Irwin
\& Saikia 2003 for examples)  we find a typical halo magnetic field
strength of $5~\mu G$, implying  ${\rm P}_B\,\sim10^{-12}$ for a unity
filling factor, or higher if the filling factor is less than unity.
The hot gas pressure in the halo is $\sim10^{-12}\rm~dyn~cm^{-2}$,
derived from the hot gas parameters listed in Table \ref{halopara}.
Although we expect variation in these quantities and filling factors
are unknown, it is reasonable to conclude that the magnetic field
will be dynamically important in any outflow model.  Parker
instabilities may also be present. Observations of the current field
configuration shows roughly vertical fields across the disk-halo
interface
 (see
T{\"ull}mann et al. 2000) although the field configuration far from
the plane has yet to be observationally determined.

The feedback energy of the young and old stellar populations can be
estimated from the IR (Heckman, Armus \& Miley 1990) and K band
(Mannucci et al. 2005) luminosity, which is
$\sim3.7\times10^{42}\rm~ergs~s^{-1}$ in total. It is not easy to
estimate the flow velocity of the hot gas, but we can adopt the
sound speed in the higher temperature hot gas as an upper limit,
which is $\sim400\rm ~km~s^{-1}$. Using this velocity, the flowing
time in the soft X-ray emitting halo is $>2.5\times10^7\rm~yr$
(adopting an extension of the soft X-ray emitting halo to be 10 kpc).
This corresponds to a total energy feedback $>3\times10^{57}\rm~ergs$,
enough to support the soft X-ray emission and the energy of
the hot gas.

\section{Summary}\label{sec:summary}

We have conducted a detailed {\sl Chandra} study of NGC~5775 and
compared the results to multi-wavelength data obtained from {\sl
Spitzer} and {\sl HST} as well as high resolution ground-based
H$\alpha$, radio continuum and \ion{H}{1} data. The main results are
as follows:

Ninety-nine discrete X-ray sources are detected in or near NGC 5775,
27 within the $D_{25}$ region, including a ULX with an absorbed
(intrinsic) 0.3-7 keV luminosity of about
$2.4~(6.5)\times10^{40}\rm~ergs~s^{-1}$.

Large-scale diffuse X-ray emission is detected to an extent nearly
10 kpc above the galactic disk in an extended halo region. There are
several prominent plume-like features detected in the halo
indicating hot gas outflows from the disk, some of which are clearly
associated with the in-disk star formation regions. In particular, a
huge shell-like feature appears in the southern halo, and has a
radius of nearly 5 kpc. The soft X-ray luminosity of this
feature is $\sim6.7\times10^{38}~\rm ergs~s^{-1}$ in the 0.3-2 keV
band. On the opposite side, a large-scale ``blob'' is detected with
a maximum projected distance $\sim25\rm~kpc$ from the galactic
plane.

Our X-ray spectral analysis reveals that the halo hot gas can
be decomposed into two characteristic components, with temperatures
of about $0.2\rm~keV$ and $0.6\rm~keV$. The total luminosity of the
halo hot gas is $\sim3.5\times10^{39}\rm~ergs~s^{-1}$.

The multi-phase halo is energetically consistent with
stellar feedback. Current star
formation with more mass loading may produce the lower temperature hot
gas component, while recent and past star formation with less mass
loading may be responsible for the higher temperature component.
The soft X-ray
shell-like feature likely represents an outflowing wind bubble
from the galactic central region. The northeast blob may be an evolved
version of such a bubble that is still confined by strong magnetic field.
We find that the magnetic
pressure is comparable to the hot gas pressure, and will therefore play an
important role in the evolution of a multi-phase halo.

We have obtained an ``X-ray-star formation'' relation for
sub-galactic scale star formation regions in this disk-wide star
forming galaxy, and this relation is consistent with that inferred
from the integrated galactic properties.

\acknowledgements

The authors wish to thank D. Calzetti and Q.-S. Gu for helpful
comments. This work is supported by NASA through the grant
NNG07AH28G, NSFC through the grants 10725312 and 10673003, and by
NASA through a grant from STScI, which is operated by AURA under
NASA contract NAS\,5-26555. This work is also supported by China
Scholarship Council.

\begin{deluxetable}{lrrrrrrrr}
  \tabletypesize{\footnotesize}
  \tablecaption{Basic Information of NGC 5775}
  \tablewidth{0pt}
  \tablehead{
  \colhead{Parameter} &
  \colhead{NGC 5775} &
    }
  \startdata
     Morphology$\rm ^a$         &  Sb(f)\\
     Center position$\rm ^a$    &  R.A. 14h53m57.6s\\
     ~~~~~(J2000)                &  Dec. +03d32m40s\\
     Inclination angle$\rm ^b$  &  $86^\circ$\\
     Distance (Mpc)$\rm ^b$     &  24.8 (1\farcs $\sim$ 0.12~kpc)\\
$M_{HI}~(10^{9}\rm~M_\odot)$$\rm ^b$  &  $9.1\pm0.6$\\
$M_T~(10^{11}\rm~M_\odot)$$\rm ^b$  &  $1.5\pm0.2$\\
$v_{max,g}\rm~(km~s^{-1})$$\rm ^c$      &  $188.78\pm6.53$\\
     $D_{25}$ (arcmin)$\rm ^a$  &  4.2\\
     $D_{SF}$ (arcsec)$\rm ^d$  &  $200\pm10$\\
Galactic foreground $N_H (10^{20}\rm~cm^{-2})^e$ & 3.48\\
     $L_{K}~(10^{10}\rm~L_\odot)$$\rm ^a$  &  1.28\\
     $L_{IR}~(10^{10}\rm~L_\odot)$$\rm ^f$  &  5.38\\
$SFR_{IR}\rm~(M_\odot~yr^{-1})^g$ & 9.1\\

\enddata
\tablecomments{References. - a. NED, $D_{25}$ is the diameter at
$I_B=25\rm~mag~arcsec^{-2}$, $L_K$ is the 2MASS K-band luminosity;
b. Irwin (1994), $M_{HI}$ is the \ion{H}{1} mass, $M_T$ is the total
mass; c. HYPERLEDA, $v_{max,g}$ is the apparent maximum rotation
velocity of gas; d. $D_{SF}$ is the star formation diameter,
averaged from H$\alpha$, radio continuum and dust emission; e.
HEASARC web tools; f. IRAS IR luminosity
$L_{IR}=5.67\times10^5D_{Mpc}^2(13.48f_{12}+5.16f_{25}+2.58f_{60}+f_{100})L_\odot$,
where $D_{Mpc}$ is the distance to the galaxy in Mpc and $f_{12}$,
$f_{25}$, $f_{60}$, and $f_{100}$ are the 12, 25, 60, and 100 $\mu$m
fluxes, respectively, in Jy; g. $SFR_{IR}$ is calculated from
$L_{IR}$ using the Kennicutt (1998) relation.}\label{basicpara}
  \end{deluxetable}
  \vfill

\begin{deluxetable}{lrrrrrrrr}
  \tabletypesize{\footnotesize}
  \tablecaption{{\sl Chandra} Source List \label{acis_source_list}}
  \tablewidth{0pt}
  \tablehead{
  \colhead{Source} &
  \colhead{CXOU Name} &
  \colhead{$\delta_x$ ($''$)} &
  \colhead{CR $({\rm~cts~ks}^{-1})$} &
  \colhead{HR} &
  \colhead{HR1} &
%  \colhead{HR2} &
  \colhead{Flag} \\
  \noalign{\smallskip}
  \colhead{(1)} &
  \colhead{(2)} &
  \colhead{(3)} &
  \colhead{(4)} &
  \colhead{(5)} &
  \colhead{(6)} &
  \colhead{(7)}
%  \colhead{(8)} &
% \colhead{(9)}
  }
  \startdata
   1 &  J145338.06+033401.3 &  1.1 &$     0.45  \pm   0.26$& --& --& B \\
   2 &  J145339.94+033418.9 &  0.6 &$     7.19  \pm   0.87$& $-0.01\pm0.10$ & $ 0.09\pm0.19$ & B, S, H \\
   3 &  J145340.60+033340.6 &  0.9 &$     0.55  \pm   0.21$& --& --& B, S, H \\
   4 &  J145341.43+033442.2 &  0.9 &$     0.25  \pm   0.13$& --& --& B, H \\
   5 &  J145342.23+033142.8 &  0.8 &$     0.58  \pm   0.14$& --& --& B, H, S \\
   6 &  J145342.49+033248.2 &  0.8 &$     0.37  \pm   0.14$& --& --& S, B \\
   7 &  J145342.77+033503.3 &  0.7 &$     1.21  \pm   0.42$& --& --& B, S \\
   8 &  J145342.81+033704.8 &  0.8 &$     3.37  \pm   0.54$& $-0.00\pm0.14$ & --& B, S, H \\
   9 &  J145343.05+033647.2 &  1.9 &$     0.37  \pm   0.21$& --& --& S \\
  10 &  J145343.78+033427.1 &  0.5 &$     2.60  \pm   0.50$& $-0.03\pm0.17$ & --& B, S, H \\
  11 &  J145344.02+032929.7 &  1.7 &$     1.51  \pm   0.78$& --& --& S \\
  12 &  J145344.69+033330.8 &  0.4 &$    21.04  \pm   0.84$& $-0.22\pm0.05$ & $ 0.19\pm0.05$ & B, S, H \\
  13 &  J145344.83+033458.9 &  0.8 &$     0.26  \pm   0.10$& --& --& H, B \\
  14 &  J145344.84+033257.5 &  0.5 &$     1.12  \pm   0.20$& --& --& B, S, H \\
  15 &  J145345.17+033348.6 &  0.5 &$     0.63  \pm   0.13$& --& --& B, S, H \\
  16 &  J145346.26+033350.4 &  0.6 &$     0.41  \pm   0.10$& --& --& B, H \\
  17 &  J145346.28+033347.1 &  0.6 &$     0.30  \pm   0.12$& --& --& B, S \\
  18 &  J145346.37+033023.0 &  1.6 &$     0.16  \pm   0.07$& --& --& H \\
  19 &  J145346.50+033459.5 &  0.5 &$     0.40  \pm   0.12$& --& --& B, H \\
  20 &  J145346.63+033257.8 &  0.4 &$     0.62  \pm   0.15$& --& --& B, S, H \\
  21 &  J145346.64+033731.2 &  1.2 &$     0.76  \pm   0.37$& --& --& S \\
  22 &  J145347.51+033341.5 &  0.5 &$     0.20  \pm   0.07$& --& --& B, H \\
  23 &  J145347.89+033404.8 &  0.4 &$     0.47  \pm   0.12$& --& --& B, S, H \\
  24 &  J145348.28+033402.3 &  0.3 &$     2.64  \pm   0.26$& $ 0.12\pm0.12$ & $ 0.78\pm0.12$ & B, S, H \\
  25 &  J145348.62+033158.8 &  0.4 &$     1.06  \pm   0.19$& --& --& B, S, H \\
  26 &  J145349.26+033137.8 &  0.6 &$     0.17  \pm   0.08$& $ 1.00\pm0.14$ & --& B, H \\
  27 &  J145349.29+033411.4 &  0.6 &$     0.16  \pm   0.06$& --& --& B, H \\
  28 &  J145350.93+032909.9 &  0.9 &$     0.35  \pm   0.11$& --& --& B, H \\
  29 &  J145350.94+033124.2 &  0.6 &$     0.17  \pm   0.07$& $ 1.00\pm0.11$ & --& B, H \\
  30 &  J145351.89+033524.2 &  0.4 &$     1.19  \pm   0.30$& --& --& B, S, H \\
  31 &  J145352.30+033317.9 &  0.4 &$     0.76  \pm   0.15$& --& --& B, S, H \\
  32 &  J145352.50+033405.8 &  0.6 &$     0.41  \pm   0.10$& --& --& B, H, S \\
  33 &  J145352.65+033140.6 &  0.3 &$     1.01  \pm   0.17$& --& --& B, S, H \\
  34 &  J145353.13+033255.4 &  0.3 &$     1.54  \pm   0.19$& $ 0.70\pm0.10$ & --& B, H, S \\
  35 &  J145353.70+033623.9 &  0.5 &$     2.59  \pm   0.37$& $-0.26\pm0.14$ & --& B, S, H \\
  36 &  J145354.13+033154.3 &  7.5 &$     0.13  \pm   0.07$& --& --& B \\
  37 &  J145354.19+033318.9 &  0.8 &$     0.29  \pm   0.09$& --& --& B, S, H \\
  38 &  J145354.30+033349.8 &  0.5 &$     0.76  \pm   0.14$& --& --& B, H, S \\
  39 &  J145354.33+032707.7 &  1.6 &$     0.25  \pm   0.10$& --& --& H \\
  40 &  J145354.36+033548.6 &  0.9 &$     0.09  \pm   0.06$& $ 1.00\pm0.16$ & --& H \\
  41 &  J145354.68+033043.2 &  0.5 &$     0.49  \pm   0.11$& --& --& B, S, H \\
  42 &  J145354.68+033345.7 &  1.5 &$     0.15  \pm   0.07$& $ 1.00\pm0.17$ & --& B \\
  43 &  J145354.74+033340.5 &  0.3 &$     1.00  \pm   0.16$& $-0.10\pm0.19$ & $ 0.86\pm0.15$ & B, S, H \\
  44 &  J145354.77+033225.4 &  0.8 &$     0.22  \pm   0.07$& --& --& H, B \\
  45 &  J145355.08+033332.5 &  1.3 &$     0.16  \pm   0.06$& --& --& H \\
  46 &  J145355.14+033245.8 &  0.5 &$     0.48  \pm   0.12$& --& --& B, S, H \\
  47 &  J145355.15+032947.9 &  0.6 &$     0.63  \pm   0.14$& --& --& S, B \\
  48 &  J145355.24+033229.1 &  0.3 &$     3.54  \pm   0.31$& $-0.15\pm0.11$ & $ 0.49\pm0.12$ & B, S, H \\
  49 &  J145355.28+032539.9 &  2.1 &$     0.76  \pm   0.37$& --& --& B, H \\
  50 &  J145355.65+033006.3 &  1.0 &$     0.12  \pm   0.06$& $ 1.00\pm0.17$ & --& H \\
  51 &  J145355.75+033328.0 &  0.2 &$     4.69  \pm   0.33$& $ 0.74\pm0.05$ & $ 1.00\pm0.03$ & B, H, S \\
  52 &  J145355.81+033518.4 &  0.5 &$     0.28  \pm   0.12$& --& --& B, H \\
  53 &  J145355.89+033430.1 &  0.5 &$     0.47  \pm   0.13$& --& --& B, S \\
  54 &  J145355.90+032913.9 &  0.6 &$     3.25  \pm   0.32$& $-0.12\pm0.12$ & $ 0.31\pm0.13$ & B, S, H \\
  55 &  J145355.92+033400.7 &  0.2 &$     7.15  \pm   0.44$& $ 0.12\pm0.07$ & $ 0.57\pm0.09$ & B, H, S \\
  56 &  J145356.25+033303.2 &  0.5 &$     0.61  \pm   0.13$& --& --& B, S, H \\
  57 &  J145356.43+032706.8 &  1.3 &$     0.62  \pm   0.15$& --& --& B, S \\
  58 &  J145356.75+033308.4 &  1.7 &$     0.23  \pm   0.08$& --& --& B \\
  59 &  J145356.79+033129.3 &  0.3 &$     2.66  \pm   0.27$& $-0.17\pm0.12$ & $ 0.68\pm0.12$ & B, S, H \\
  60 &  J145356.85+033240.8 &  0.8 &$     0.32  \pm   0.11$& --& --& B, H \\
  61 &  J145356.89+033257.5 &  0.4 &$     0.94  \pm   0.16$& $-0.01\pm0.19$ & --& B, S, H \\
  62 &  J145357.15+033251.3 &  0.5 &$     0.58  \pm   0.13$& --& --& B, S, H \\
  63 &  J145357.35+033243.0 &  0.5 &$     0.35  \pm   0.10$& $ 1.00\pm0.13$ & --& H, B \\
  64 &  J145357.42+033226.2 &  0.7 &$     0.23  \pm   0.08$& --& --& B, H \\
  65 &  J145357.62+033241.8 &  0.5 &$     0.51  \pm   0.12$& --& --& B, H, S \\
  66 &  J145357.66+033236.6 &  0.8 &$     0.33  \pm   0.09$& --& --& H \\
  67 &  J145357.66+033030.2 &  0.5 &$     0.53  \pm   0.11$& $ 0.91\pm0.10$ & --& B, H \\
  68 &  J145357.71+033240.4 &  0.8 &$     0.19  \pm   0.07$& --& --& B \\
  69 &  J145357.76+033250.3 &  0.9 &$     0.13  \pm   0.06$& --& --& H \\
  70 &  J145357.89+033238.3 &  0.6 &$     0.36  \pm   0.10$& $ 0.92\pm0.14$ & --& B, H \\
  71 &  J145357.92+032815.3 &  1.0 &$     0.49  \pm   0.13$& --& --& B, H \\
  72 &  J145357.97+033217.3 &  1.0 &$     0.21  \pm   0.08$& --& --& B \\
  73 &  J145358.63+032744.9 &  1.1 &$     0.70  \pm   0.16$& --& --& S, B \\
  74 &  J145358.66+033506.4 &  0.4 &$     0.28  \pm   0.10$& --& --& B, S \\
  75 &  J145358.81+033303.5 &  1.2 &$     0.18  \pm   0.07$& --& --& S, B \\
  76 &  J145358.89+033216.7 &  0.2 &$    27.98  \pm   0.80$& $ 0.88\pm0.02$ & $ 0.91\pm0.07$ & B, H, S \\
  77 &  J145359.44+033157.2 &  0.3 &$     6.51  \pm   0.40$& $ 0.39\pm0.07$ & $ 0.72\pm0.10$ & H, B, S \\
  78 &  J145359.47+033147.8 &  0.4 &$     0.71  \pm   0.13$& $ 0.44\pm0.19$ & --& B, H, S \\
  79 &  J145359.61+033138.9 &  1.0 &$     0.16  \pm   0.07$& --& --& H \\
  80 &  J145359.75+033140.2 &  0.3 &$     0.74  \pm   0.14$& --& --& B, H, S \\
  81 &  J145400.15+033131.2 &  0.4 &$     0.36  \pm   0.11$& --& --& S, B \\
  82 &  J145400.81+033130.9 &  0.4 &$     0.96  \pm   0.15$& $ 1.00\pm0.03$ & --& H, B \\
  83 &  J145400.95+033129.4 &  0.3 &$     2.30  \pm   0.24$& $ 0.64\pm0.09$ & --& B, H, S \\
  84 &  J145400.96+033133.1 &  0.3 &$     2.95  \pm   0.27$& $ 0.86\pm0.05$ & --& B, H, S \\
  85 &  J145401.47+033753.3 &  0.8 &$     3.46  \pm   0.63$& $-0.30\pm0.15$ & $-0.14\pm0.20$ & B, S, H \\
  86 &  J145402.59+033014.7 &  0.7 &$     0.36  \pm   0.11$& --& --& B, S \\
  87 &  J145402.85+033917.9 &  1.3 &$     0.74  \pm   0.27$& --& --& B, S \\
  88 &  J145402.96+033210.1 &  0.4 &$     0.51  \pm   0.12$& --& --& B, S, H \\
  89 &  J145403.51+033842.8 &  1.7 &$     0.26  \pm   0.14$& --& --& H \\
  90 &  J145404.84+033422.3 &  0.5 &$     0.23  \pm   0.08$& --& --& B, H \\
  91 &  J145404.88+032722.4 &  1.4 &$     0.37  \pm   0.12$& --& --& B \\
  92 &  J145405.71+033126.4 &  0.6 &$     0.49  \pm   0.14$& --& --& B, S, H \\
  93 &  J145405.77+033028.7 &  0.6 &$     0.38  \pm   0.10$& --& --& B, S, H \\
  94 &  J145406.56+033327.0 &  0.7 &$     0.17  \pm   0.07$& --& --& B, H \\
  95 &  J145407.95+033102.1 &  0.7 &$     0.50  \pm   0.12$& --& --& B, S \\
  96 &  J145410.14+032734.2 &  1.3 &$     1.01  \pm   0.21$& --& --& B, S \\
  97 &  J145411.08+033016.1 &  1.1 &$     0.37  \pm   0.12$& --& --& B, S \\
  98 &  J145413.24+033244.9 &  0.6 &$     5.62  \pm   0.45$& $-0.12\pm0.10$ & $ 0.35\pm0.11$ & B, S, H \\
  99 &  J145414.09+033101.3 &  0.8 &$     1.04  \pm   0.19$& --& --& B, S, H \\
\enddata
\tablecomments{The definition of the bands:
0.3--0.7 (S1), 0.7--1.5 (S2), 1.5--3 (H1), and 3--7~keV (H2). % for ACIS-S
%0.5--1 (S1), 1--2 (S2), 2--4 (H1), and 4--8~keV  (H2). % for ACIS-I
%1--2.5 (S1), 2.5--4 (S2), 4--6 (H1), and 6--9~keV  (H2). % for ACIS-I_low
In addition, S=S1+S2, H=H1+H2, and B=S+H.
 Column (1): Generic source number. (2):
{\sl Chandra} X-ray Observatory (unregistered) source name,
following the {\sl Chandra} naming convention and the IAU
Recommendation for Nomenclature (e.g.,
http://cdsweb.u-strasbg.fr/iau-spec.html). (3): Position uncertainty
(1$\sigma$) calculated from the maximum likelihood centroiding and
an approximate off-axis angle ($r$) dependent systematic error
$0\farcs2+1\farcs4(r/8^\prime)^2$ (an approximation to Fig.~4 of
Feigelson, E., et al. 2002), which are added in quadrature.
%\bibitem[Feigelson et al. (2002)]{fei02} Feigelson, E., et al. 2002, ApJ, 574, 258
(4): On-axis source broad-band count rate --- the sum of the
exposure-corrected count rates in the four bands. (5-6): The
hardness ratios defined as
%${\rm HR}=({\rm H1-S})/({\rm H1+S})$, and ${\rm HR1}=({\rm H2-H1})/{\rm H}$, %dhrch=2
${\rm HR}=({\rm H-S2})/({\rm H+S2})$, and ${\rm HR1}=({\rm S2-S1})/{\rm S}$, %dhrch=0
listed only for values with uncertainties less than 0.2. (7): The
label ``B'', ``S'', or ``H'' mark the band in which a source is
detected with the most accurate position that is adopted in Column
(2). This table will be published online only.
%The label ``v'' denotes that a source is a variable.
}\label{detpointsrc}
  \end{deluxetable}
  \vfill

\begin{deluxetable}{lrrrrrrrr}
  \tablecaption{Spectral Fitting Parameters}
  \tabletypesize{\tiny}
  \tablewidth{0pt}
  \tablehead{
  \colhead{Region} & Model & $N_H$ & $T_{Low}$ & $EM_{Low}$ & $T_{High}$ & $EM_{High}$ & $\Gamma$ & $\chi^2/d.o.f.$ \\
  \colhead{} & & $(10^{20}\rm~cm^{-2})$ & (keV) & ($\rm cm^{-6}kpc^3$) & (keV) & ($\rm cm^{-6}kpc^3$) & & \\
  \colhead{(1)} & (2) & (3) & (4) & (5) & (6) & (7) & (8) & (9) \\

    }
\startdata
ULX & PL & $286_{-37}^{+50}$ & - & - & - & - & $1.82_{-0.22}^{+0.31}$ & 50.6/44 \\
Disk & Mekal+PL & $53.9_{-3.9}^{+10.5}$ & $0.19_{-0.01}^{+0.01}$ & 1.31 & - & - & 1.34 (fixed) & 92.9/74 \\
Halo & Mekal+Mekal & 3.48 (fixed) & 0.17 ($<0.19$) & 0.047 & $0.57_{-0.15}^{+0.15}$ & 0.017 & -& 70.4/69
\enddata

\tablecomments{(1) Regions from which spectra were extracted; (2) Mekal =
XSPEC MEKAL thermal plasma model, PL = Power Law; (3) Hydrogen
absorption column density; (4)(6) Temperatures of the thermal
components; (5)(7) Emission measure of the thermal components; (8)
Photon index of the PL component; (9) Statistics of the spectral
fit. Note that the last column is calculated with the sky
background, while the spectra shown in Fig. \ref{spectra} is after
sky background subtraction.}\label{specfitpara}
  \end{deluxetable}
  \vfill

\begin{deluxetable}{lrrrrrrrr}
  \tabletypesize{\footnotesize}
  \tablecaption{Hot Halo Gas Parameters}
  \tablewidth{0pt}
  \tablehead{
  \colhead{Parameters} & Low T Component & High T Component \\
    }
\startdata
Volume Filling Factor & 0.2 & 0.8 \\
Number Density ($10^{-3}~\rm cm^{-3}$) & 4.5 & 1.3 \\
Mass ($10^8~\rm M_\odot$) & 2.4 & 2.9 \\
Thermal Energy ($10^{56}\rm~ergs$) & 1.4 & 5.5 \\
Cooling Time Scale ($10^9\rm~yr$) & 2.0 & 13 \\
\enddata

\tablecomments{These quantities are derived from the parameters
listed in Table \ref{specfitpara}, assuming that the filling factor
of the hot gas is $f_{high}$ ($f_{low}$) for the high (low)
temperature component with $f_{high}+f_{low}\sim1$, and that the hot
gas is located in a cylinder with a diameter of
$D_{25}$.}\label{halopara}
  \end{deluxetable}
  \vfill

\end{document}